\documentclass[]{article}

\usepackage{latexsym}

\newcommand{\go}{\omega}
\newcommand{\sbs}{\subseteq} 
\newcommand{\eqv}{\Leftrightarrow} 
\newcommand{\dam}{\Diamond} \newcommand{\sat}{{\models}}
\newcommand{\dar}{{\downarrow}}

\newcommand{\sk}{{\sf K}}\newcommand{\sa}{{\sf A}}
\newcommand{\sfl}{{\sf L}}

 \newcommand{\cl}{{\cal L}}
 
\newcommand{\cm}{{\cal M}} \newcommand{\cp}{{\cal P}}
\newcommand{\cf}{{\cal F}} \newcommand{\cg}{{\cal G}}
\newcommand{\cn}{{\cal N}} \newcommand{\co}{{\cal O}}
\newcommand{\clc}{{\cal C}}

\newcommand{\fsk}{{\hbox{\footnotesize\sk}}}

\newcommand{\rem}[4]{\downarrow #1-\bigcup^{#3}_{#2}\downarrow #4}
\newcommand{\Rem}[2]{{\sf Rem}^{#1}#2}

\newcommand{\msat}{\sat_{\cm}}
\newcommand{\csat}{\sat_{\clc}}
\newcommand{\notcsat}{\not\sat_{\clc}}
\newcommand{\MPT}{\vdash_{\bf\scriptstyle MPT}}
\newcommand{\bland}{\bigwedge}

\newcommand{\todam}{\stackrel{\dam}{\to}}
\newcommand{\tol}{\stackrel{{\hbox{\footnotesize\sfl}}}{\to}}
\newcommand{\lmean}{[\![}
\newcommand{\rmean}{]\!]}

\newtheorem{theorem}{\bf Theorem}
\newtheorem{lemma}[theorem]{\bf Lemma}
\newtheorem{prop}[theorem]{\bf Proposition}
\newtheorem{corol}[theorem]{\bf Corollary}

\newenvironment{dfn}{\par\medskip\addtocounter{theorem}{1}%
  \noindent{\bf Definition \arabic{theorem}}\quad}{\medskip}

\newcommand{\qed}{\vrule height5pt width3pt depth0pt}

\newenvironment{prf}{\noindent {\sc Proof.}}{
{\nobreak\hfill \qed \par \medbreak}}

\title{Knowledge on Treelike Spaces}

\author{ Konstantinos Georgatos\\
Dipartimento di Informatica e Sistemistica\\
Universit\`{a} di Roma ``La Sapienza''\\
 Via Salaria 113\\
I-0019 Roma \\
Italy\\
e-mail:  {\tt geo@dis.uniroma1.it}. }

\date{April 8, 1995}

\begin{document}

\maketitle


\begin{abstract}
This paper presents a bimodal logic for reasoning about knowledge during
knowledge acquisition. One of the modalities represents (effort during)
non-deterministic time and the other represents knowledge. The semantics of this
logic are tree-like spaces which are a generalization of semantics used for
modeling branching time and historical necessity. A finite system of axiom
schemes is shown to be canonically complete for the formentioned spaces. A
characterization of the satisfaction relation implies the small model property
and decidability for this system.
\end{abstract}

\section{Introduction}
\label{sec:intro}

The notion of possible world dominates  the literature in modal logic, via
Kripke models, as well as in any logic dealing with the epistemic state of a
reasoner. The heart of this popularity lies in the identification of an
intentional state through common properties of extensional objects. Apart from
 genuine problems such as logical omniscience this representation suffers from,
it is limited in a static description of the reasoner's epistemic state. The
``logic of knowing'' is not only embodied in the representation of knowledge but
also in the way knowledge is acquired. We do not refer to temporal properties but
rather to methodology (though both can be intertwined).

 Recently a family of logics was introduced
(\cite{MP},\cite{KG1},\cite{KGT},\cite{DMP}) with the intention to fill this
void.  It succeeds in doing so by attaching familiar mathematical structures
such as spaces of subsets, topologies and complete lattices of subsets
corresponding to a natural knowledge acquisition. This paper extends this work by
introducing a bimodal logic belonging to the same family of logics and
establishes a correspondence between a particular epistemic process of knowledge
acquisition with a space of subsets forming a tree (treelike space).

In our framework the view of a reasoner will be represented by a set of
possible worlds. Each of these worlds represents an alternative state
compatible with the reasoner's knowledge of actual state.  This
treatment of knowledge agrees with the traditional one
(\cite{HI}, \cite{HAM}, \cite{PR}, \cite{CM})
expressed in a variety of contexts (artificial intelligence, distributed
processes, economics, etc).

  We are interested in formulating a basic logical framework for reasoning about
a resource-conscious acquiring of knowledge. Such a framework can be applied to
many settings such as the ones involving time, computation, physical experiments
or observations. In these settings an (discrete or continuous) increase of
information available to us takes place and results in an increase of our
knowledge. How could this simple idea  be embodied in the formentioned
semantical framework? An increase of knowledge can be represented with a
restriction of the knower's view, i.e. of the equivalence class of the
alternative worlds. This restriction is nondeterministic (we do not know what
kind of additional information will be available to us, if at all) but not
arbitrary: it will always contain the actual state of the knower, i.e. it is a
{\it neighborhood restriction} of the actual state. In this way, set-theoretic
considerations come in.

\begin{figure}
\centering
\setlength{\unitlength}{0.012500in}%
\begin{picture}(240,191)(200,480)
\thinlines
\put(240,555){\vector( 1, 0){0}}
\put(240,539){\oval( 22, 32)[tl]}
\put(247,539){\oval( 36, 36)[bl]}
\put(247,539){\oval( 34, 36)[br]}
\put(263.7,550){\line( 0,-1){11}}
\put(315,640){\vector(-3,-4){ 45}}
\put(325,640){\vector( 3,-4){ 45}}
\put(270,550){\vector( 2,-3){ 32}}
\put(365,550){\vector(-2,-3){ 32}}
\put(375,550){\vector( 2,-3){ 32}}
\put(290,655){$\{q_1,q_2,q_3,q_4\}$}
\put(245,560){$\{q_1,q_2\}$}
\put(355,560){$\{q_3,q_4\}$}
\put(440,525){$Q_2$}
\put(440,610){$Q_1$}
\put(400,485){$\{q_4\}$}
\put(320,485){$\{q_3\}$}
\put(300,485){$\emptyset$}
\put(270,610){{\sf yes}}
\put(330,530){{\sf yes}}
\put(360,610){{\sf no}}
\put(400,530){{\sf no}}
\put(200,530){{\sf yes}}
\put(290,530){{\sf no}}
\end{picture}
\caption{A knowledge acquisition tree.}
\label{fig:tree}
\end{figure}
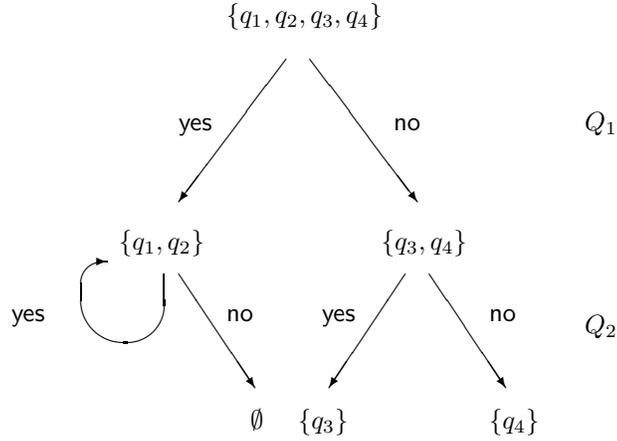

A discrete version of our epistemic framework can arise in
scientific experiments or tests. We acquire knowledge by ``a step-by-step''
process, each step being an experiment or test. The outcome of such an
experiment or test is unknown to us beforehand, but after being known it
restricts our attention to a smaller set of possibilities. A sequence of
experiments, tests, or actions comprises a {\em strategy of knowledge
acquisition}. This model is in many respects similar to Hintikka's ``oracle''
(see~\cite{HI86}). In Hintikka's model the ``inquirer'' asks a series of
questions $Q_1, Q_2, \ldots, Q_n,\ldots $ to an external information source,
called ``oracle'' (can be thought as a knowledge base). The oracle answers yes
or no and the inquirer increases his or her knowledge by this piece
of additional
evidence. At any point of this process the inquirer follows a branch of a tree
determined by the possible answers to his or her series of questions. Such an
interrogative model is recognized by Gadamer (\cite{GAD}) as an important
part of the epistemic process. Consider the following example:
\medskip

\noindent {\em Example:\/}
Suppose that our view, the set of possible worlds, is $\{ q_1, q_2, q_3, q_4\}$
and our query consists of two questions $Q_1$, $Q_2$, in that order. The answer
to $Q_1$ is {\sf yes} in $q_1$, $q_2$ and {\sf no} in $q_3$, $q_4$. The answer to
$Q_2$ is {\sf yes} in $q_1$, $q_2$, $q_3$ and {\sf no} in $q_4$. Then the possible
sequences of knowledge states comprise a tree of subsets as shown in
Figure~\ref{fig:tree}. The space
of subsets labeling the nodes of the  tree will be called a
{\em treelike space}.

The above example shows a transition from the symbolic description of the
epistemic process to a description in spatial terms. Instead of going down a
proof tree, the one which entails the desired formulae, we intersect nodes of a
tree labeled by subsets of a space. This transition is direct; it enables us
to think in {\em geometric} terms.

Now consider the following example:
\medskip

\noindent {\em Example:\/}
 Suppose that a machine emits a stream of binary digits representing
the output of a recursive function $f$. After time $t_1$  the machine emitted
the stream $111$. The only information we have about the function being computed
at this time on the basis of this (finite) observation is that
$$f(1)=f(2)=f(3)=1.$$
As far as our knowledge concerns, $f$ is indistinguishable from the constant
function ${\bf 1}$, where ${\bf 1}(n)=1$ for all $n$. After some additional time
$t_2$, i.e. spending more time and resources, $0$ might appear and thus we could
be able to distinguish $f$ from $\bf 1$.
In any case, each binary stream will be an initial segment of $f$ and this
initial segment is a neighborhood of $f$. In this way, we can acquire better
knowledge of the function the machine computes. The space of finite binary
streams is a structure which models computation.  The sets of binary streams
under the initial segment ordering is an example of a treelike space.

The above example shows how the same epistemic process appears during
observations of programs. Here possible worlds correspond to (total)
computations and our view to observations. We can apply  the same
spatial reasoning to programs through the following correspondence:

\begin{center}
\begin{tabular}{rcccl}
Knowledge states & $=$ & Sets & $=$ & Observations \\
\medskip

Possible worlds & $=$ & Points & $=$ & Computations.
\end{tabular}
\end{center}

Therefore a common idea lies behind the knowledge-theoretic, spatial and
computational framework. The connection between the last two is not new.
Here is how this epistemic framework ties with previous work on establishing
links between spatial reasoning and reasoning about programs.

We use two
modalities  $\sk$ for knowledge and $\Box$
for effort, i.e. spending of resources. Consider the formula
$$
A\to\dam\sk A,
$$
where $A$ is an atomic predicate and $\dam$ is the dual of the $\Box$,
i.e. $\dam\equiv\neg\Box\neg$. It will be clear after the presentation of
semantics in Section~\ref{sec:lang} that if the above
formula is valid, then the set which
$A$ represents is an open set of the topology generated by the subsets of the
treelike space as a basis. Under the reading of $\dam$ as ``possible'' and $\sk$
as ``is known'', the above formula says that
\begin{quote}
``if $A$ is true then it is possible for $A$ to be known'',
\end{quote}
i.e. $A$ is {\em affirmative}. Vickers defines similarly an
affirmative assertion  in~\cite{V}
\begin{quote}
``an assertion is affirmative iff it is true precisely in the circumstances when
it can be affirmed.''
\end{quote}
Affirmative and refutative assertions are closed under infinite disjunctions and
conjunctions, respectively. Smyth
in~\cite{SMY83} observed first these properties in semi-decidable properties.
Semi-decidable properties are those properties whose truth set is r.e. and are a
particular kind of affirmative assertions. In fact, changing our power of
affirming or computing we get another class of properties with a similar
knowledge-theoretic character. For example, using polynomial algorithms
affirmative assertions become polynomially semi-decidable, i.e. NP
properties. If an object has this property then it is possible to know it with a
polynomial algorithm even though it is not true we know it now.

Our approach has an independent theoretical interest. A new family of
Kripke frames, called {\it subset frames}, arises. These are Kripke frames which
are equivalent to sets of subsets. In particular, we have identified those which
are equivalent to (complete) lattices of subsets and topologies
(see~\cite{KGT}). In this paper, we shall identify those which correspond to the
above interrogative model, called {\em treelike spaces}. Treelike spaces
have a particular interest; they correspond to an indeterminist's theory of time
called {\em Ockhamism} (see~\cite{PRI}), which gives rise to branching time.
We refer the reader to section~\ref{sec:lang} for a detailed
discussion.

A family of logics for knowledge and time is studied in~\cite{HAV} and various
complexity results are established. However, the framework of the above logics is
restricted to distributed systems and their interpretation differs significantly
from ours.

Interpreting the knowledge modal operator as a universal quantifier we present
a novel way of  understanding the meaning of quantifiers in varying
(ordered) domains  (see section~\ref{sec:mps} for a relevant
discussion). This is one of the main difficulties in formulating a  meaningful
first-order system for modal logic (see~\cite{FIT93} for a  discussion).

The language and semantics of our logical framework is presented in
Section~\ref{ch:systems}. In the same section, we present two systems  which
belong to the same family of logics, studied in~\cite{MP}, \cite{KGT}
and~\cite{KG1}.   In Section~\ref{ch:mpt}, we present an axiomatization, called
$\bf MPT$, for our semantics and we prove completeness, small
model property, and decidability.

A preliminary version of this paper has appeared in~\cite{KG94}.

\section{Two Systems: $\bf MP$ and $\bf MP^*$}
\label{ch:systems}

\subsection{Language and Semantics}
\label{sec:lang}

We follow the notation of \cite{MP}.

We construct a bimodal propositional modal logic.
Formally, we start with a countable set $\sa$ of {\em atomic
formulae}, then the {\em language} $\cl$ is the
least set such that $\sa\sbs\cl$ and closed under the following rules:

$$\frac{\phi,\psi\in\cl}{\phi\land\psi\in\cl}\qquad
\frac{\phi\in\cl}{\neg\phi,\Box\phi,\sk\phi\in\cl}$$

We abbreviate, as usual, $\phi\land\neg\phi$ with $\bot$ and $\neg\bot$
with $\top$.
The language $\cl$ can be interpreted inside any spatial context as follows.

\begin{dfn}
Let $X$ be a set and $\co$ a subset of the powerset of $X$, i.e.
$\co\sbs\cp(X)$ such that $X\in\co$.
We call the pair $\langle X,\co \/\rangle$ a {\em subset space}.
A {\em model} is a triple $\langle X,\co,i\/\rangle$, where
$\langle X,\co \rangle$ is a subset space and $i$ a map from $\sa$
to $\cp(X)$ with $i(\top)=X$ and $i(\bot)=\emptyset$ called {\em
initial interpretation}.
\end{dfn}

We denote the set
$\{ (x,U) : U\in\co , \hbox{ and } x\in U\}\sbs X\times\co$
with $X\dot{\times}\co$. For each $U\in\co$ let $\dar U$  be the lower closed
set generated by $U$ in the partial order $(\co,\sbs)$, i.e. the set
$\{ V : V\in\co \hbox{ and } V\sbs U \}$.

\begin{dfn}
The {\em satisfaction relation} $\msat$, where $\cm$ is the model
$\langle X,\co,i\/\rangle$,
is a subset of $(X\dot{\times}\co)\times\cl$ defined recursively  by
(we write $x,U\msat\phi$ instead of $((x,U),\phi)\in\msat$)
$$
\begin{array}{ll}
x,U\msat A & \hbox{iff}\quad x\in i(A),\hbox{ where }A\in\sa \medskip
\\ x,U\msat \phi\land\psi & \hbox{if}\quad x,U\msat\phi \hbox{ and }
                                           x,U\msat \psi\medskip  \\
x,U\msat \neg\phi & \hbox{if}\quad x,U \not\msat\phi        \medskip
\\ x,U\msat \sk\phi & \hbox{if}\quad\hbox{for all } y\in U,\quad
                                           y,U\msat\phi \medskip  \\
x,U\msat \Box\phi & \hbox{if} \quad\hbox{for all } V\in\dar U \hbox{
such that } x\in V, \quad x,V\msat\phi.
\end{array}
$$
If $x,U\msat\phi$, for all $(x,U)$ belonging to $ X\dot{\times}\co$,
then $\phi$ is {\em valid} in $\cm$, denoted by $\cm\sat\phi$.
\end{dfn}

The case for atomic formulae shows that we deal with analytic sentences, i.e.
sentences which do not change their truth value. If a formula $\Box\phi$ does not
contain $\sk$ then it has the same interpretation as $\phi$.   This has also the
consequence that the universal substitution rule does not hold. Thus, time does
not  affect the semantic value of sentences but rather the knowledge we have of
them. This difference makes the
$\Box$ modality not collapsing to a temporal modality but being closer to
necessity.

We abbreviate $\neg\Box\neg\phi$ and $\neg\sk\neg\phi$ with
$\dam\phi$ and $\sfl\phi$ respectively.
We have that
$$
\begin{array}{ll}
x,U\msat \sfl\phi & \hbox{if there exists } y\in U \hbox{ such that }
                                             y,U\msat\phi \medskip \\
x,U\msat \dam\phi & \hbox{if there exists } V\in\co \hbox{ such that }
V\sbs U,\  x\in V,\hbox { and } x,V\msat\phi.
\end{array}
$$

\begin{dfn}\label{dfn:treelike}
A {\it treelike  space} is a subset space $\langle X,\co \/\rangle$ where
for all $U,V\in\co$, either $U\sbs V$, or $V\sbs U$, or $U\cap V=\emptyset$.
A model induced by a tree space will be called a {\em treelike model}.
\end{dfn}

It is clear that in the countable case the set of subsets of a treelike space
forms a tree under the subset ordering.\medskip

\noindent {\em Example: \/} Let
$$X=\{ f \mid f \hbox{ recursive }\}.$$
Now, let
$$[a_1,a_2,\ldots,a_n]=\{ f \mid f(k)=a_k, \hbox{ for } k=1,2,\ldots,n \}\sbs
X,$$
where $a_1,a_2,\ldots,a_n$ are natural numbers,
and
$$\co=\{ [a_1,a_2,\ldots,a_n] \mid n=1,2,\dots \} \cup \{X\}.$$
Then it is easily verified, using definition~\ref{dfn:treelike}, that
$\langle X,\co \/\rangle$ is a treelike space.

Now let $\bf 1$ be a predicate with
$$i({\bf 1})=\{f\mid \hbox{ there exists } n
\hbox{ such that for all } m>n, f(n)=1\}.$$ Then the formula
$$\Box\sfl {\bf 1}$$
which translates to ``it will never be known that $0$ appears infinitely often",
is valid in the treelike model $\langle X,\co,i \/\rangle$. This comes with no
surprise, since the knowledge of ``infinitely often" requires an infinite amount
of resources. This formula is an example of a refutative assertion (see
introduction).\medskip

Treelike spaces get their name from treelike frames (see~\cite{PRI}). A {\em
treelike frame} is a pair $\langle T,<\rangle$, where $T$ is a nonempty set and
$<$ is a transitive ordering on $T$ such that if $t_1<t$ and $t_2<t$ then
either $t_1=t$ or $t_1<t_2$ or $t_2<t_1$. Treelike frames have appeared as
semantics for the Ockhamist's concept of non-deterministic time
 and been used for treating historical necessity and conditionals
(see~\cite{THO}  and \cite{VAF}). The validity on these frames is called {\em
Ockhamist  validity}. A treelike space is a special form of a treelike frame
where the temporal instants of the frame are labeled by subsets of a space and
whenever instants are incomparable the respective subsets are disjoint. It can
be easily seen that the ordering among subsets is a treelike frame. The
similarities do not end here. Let $\langle T,<\rangle$ be a treelike frame and,
for each $t\in T$, $B_t$ the set of maximal linear ordered subsets of $T$
containing $t$, i.e. the branches intersecting $t$. Then
$\{B_t\}_{t\in T}$ is a treelike space. The difference lies on the
interpretation of atomic formulae. We interpret atomic formulae on branches
while an Ockhamist assignment interprets atomic formulae on temporal instances.
This bring up another dimension of our logic. Our logic is not conservative
over a logic which interprets $\Box$ as $\sf F$ (the ``future'' modality) for
if $\phi$ contains no occurrences of $\sk$ then $\Box\phi$ is valid in a
treelike space exactly when $\phi$ is. We adopt the indeterminist's view of
necessity (knowledge). Although $\phi$ may be true in our world, $\sk\phi$ may be
false. This is because there is no special world in our view which deserves to be
called {\em actual}.  Setting apart Ockhamist validity, treelike spaces  are
more general than treelike frames (and their derivative $T\times W$ frames) due
to the fact that we do not assume an overall temporal ordering. In this sense
treelike spaces are closer to a more general structure, first introduced by
Kamp and subsequently called {\em Kamp frames}, where worlds do not participate
in the same temporal structure (for definition and discussion see~\cite{THO}).
In fact, it is easily seen that treelike spaces are equivalent to {\it Ockhamist
frames} introduced by Zanardo in~\cite{ZAN85} for the completeness of strong
Ockhamist validity. At any rate, our work seems to have more than superficial links
with work in historical necessity and questions such as what the connections
between the two notions of validity are should be the subject of a more systematic
investigation.

\subsection{$\bf MP$ and $\bf MP^*$}
\label{sec:mps}

We saw that the semantics of the bimodal language is interpreted in any pair
$\langle X,\co \/\rangle$. What happens when we allow $\co$ to be any class of
sets of subsets? If $\co$ is an arbitrary set of subsets then the system $\bf
MP$ is complete for such subset spaces.  The
axiom system $\bf MP$ consists of axiom schemes \ref{ax:prop}
through~\ref{ax:boxk} and rules of  Table~\ref{table:mp}
(see page~\pageref{table:mp}) and appeared first in \cite{MP}.

The following was proved in \cite{MP}.
\begin{theorem}
The axioms and rules of $\bf MP$ are sound and complete with respect to subset
spaces.
\end{theorem}

If $\co$ is a complete lattice under set-theoretic union and intersection then
the system $\bf MP^*$ is canonically complete for this class of subset spaces.
 The axiom system $\bf MP^*$ consists of the axiom schemes and
rules of $\bf MP$ plus the following two additional axiom schemes:

$$\dam\Box\phi\to\Box\dam\phi$$
and
$$\dam(\sk\phi\land\psi)
       \land\sfl\dam(\sk\phi\land\chi)
       \to\dam(\sk\dam\phi\land\dam\psi\land\sfl\dam\chi).$$

The first axiom is a well-known formula which characterizes
{\em incestual} frames, i.e. if two points $\beta$ and $\gamma$ in a
frame can be accessed
by a common point $\alpha$ then there is a point $\delta$ which
can be accessed by both $\beta$ and $\gamma$.
The second characterizes union.

The following was proved in~\cite{KGT}.
\begin{theorem}
The axioms and rules of $\bf MP^*$ are sound and canonically complete with
respect to subset spaces, which are complete lattices.
\end{theorem}

The proof of the above theorem was later shortened and improved through an
elegant embedding of $\bf S4$ (and therefore intuitionistic logic via the
G\"{o}del translation) by Dabrowski, Moss and Parikh in~\cite{DMP}. This
translation reveals that truth in intuitionistic logic coincides with
``possibility of knowing'' in our system. It also reveals a connection with
another line of work, that of Fischer Servi. In~\cite{FIS80} and~\cite{FIS84} the
semantics and syntax of the family $*$-IC of intuitionistic modal logics is
studied. This family is is naturally embedded via the G\"{o}del translation to
the family ($\bf S4$-$*$) of bimodal logics, where $\bf S4$ is always one of the
coordinates (like in our case). However, the semantics called {\it double model
structures} (birelational modal frames) deviate from our space theoretic
framework; a fact that declares itself on the presence of different
connecting axioms, i.e. axioms involving both modalities.

\section{The system $\bf MPT$ }\label{ch:mpt}

We add the axioms~\ref{ax:tree} and~\ref{ax:ntree} to form the system $\bf
MPT$ for the purpose
of  axiomatizing treelike spaces.

\begin{table}[!t]
\begin{center}
\vspace{.3in}
\noindent{\bf Axioms}

\begin{enumerate}

\item
All propositional tautologies \label{ax:prop}

\item
$(A\to\Box A) \land (\neg A\to\Box\neg A)$, for
$A\in\sa$\label{ax:atom}

\item
$\Box(\phi\to\psi)\to(\Box\phi\to\Box\psi)$\label{ax:normb}

\item
$\Box\phi\to\phi$

\item
$\Box\phi\to\Box\Box\phi$\label{ax:s4}

\item
$\sk(\phi\to\psi)\to(\sk\phi\to\sk\psi)$\label{ax:normk}

\item
$\sk\phi\to\phi$

\item
$\sk\phi\to\sk\sk\phi$

\item
$\phi\to\sk\sfl\phi$\label{ax:s5}

\item
$\sk\Box\phi\to\Box\sk\phi$\label{ax:boxk}



\item
$\Box(\Box\phi\to\psi)\lor\Box(\Box\psi\to\phi)$\label{ax:tree}

\item
$\Box\sk\phi \land \sk ( \Box\phi\to\Box\psi ) \to \Box\sk ( \Box\phi
\to\Box\psi)$ \label{ax:ntree}

\end{enumerate}

\noindent{\bf Rules}

$$\frac{\phi\to\psi,\phi}{\psi}\ \hbox{\footnotesize MP}$$
$$\frac{\phi}{\fsk\phi}\ \hbox{\footnotesize
\footnotesize \sk-Necessitation} \qquad
  \frac{\phi}{\Box\phi}\ \hbox{\footnotesize $\Box$-Necessitation}$$
\medskip
\end{center}
\caption{\label{table:mp} Axioms and Rules of $\bf MPT$.}
\end{table}

A word about the axioms (most of the following
facts can be
found in any introductory book about modal logic, e.g. \cite{CH} or \cite{GL}.)
Axiom~\ref{ax:atom}  expresses the fact that the truth of atomic formulae
is independent of the choice of subset and depends only on the choice of point.
Axioms \ref{ax:normb} through
\ref{ax:s4} and Axioms \ref{ax:normk} through \ref{ax:s5} are used to axiomatize
the normal modal logics {\bf S4} and {\bf S5} respectively. The former group of
axioms expresses the fact that the passage from one subset to its restriction
 is done in a constructive way, as actually happens in an increase of
information or a spending of resources (the classical
interpretation of necessity in intuitionistic logic is axiomatized in the
same way). The latter group is generally used for axiomatizing logics of
knowledge.

Axiom~\ref{ax:boxk} expresses the fact that if a formula holds in arbitrary
subsets is going to hold as well in the ones which are neighborhoods of a
point. The converse of this axiom is not sound.


Axiom~{\ref{ax:tree}} is a well-known axiom which characterizes reflexive,
transitive and {\em connected} frames, i.e. if two points $\beta$ and $\gamma$ in a
frame can be accessed
by a common point $\alpha$ then either  $\beta$ accesses
$\gamma$ or $\gamma$ accesses $\beta$ (or both).

Soundness of Axioms \ref{ax:prop}~through~\ref{ax:boxk} has already been
established for arbitrary subset spaces (see~\cite{MP}). The soundness of
Axiom~\ref{ax:tree} is easy to see, since the {\em
subset frame} (see \cite{KGT}), i.e. the birelational modal frame, of a tree model
is  connected.

\begin{prop}
The axiom~\ref{ax:ntree} is sound.
\end{prop}

\begin{prf}
We shall show soundness for the equivalent formula
$$\Box\sk\phi \land \dam\sfl ( \psi\land\Box\phi ) \to \sfl ( \dam\psi \land
\Box\phi).$$
Let $x,U\sat \Box\sk\phi \land \dam\sfl ( \psi\land\Box\phi )$. Then there
exists $V\sbs U$ such that $x,V\sat \sfl ( \psi\land\Box\phi )$. This implies
that
there exists $y\in V$ such that $y,V \sat \psi\land\Box\phi $.
Now, observe that $y,U\sat \Box\phi$. For, if $W\sbs U$ and $y\in W$ then there
are two cases. Either $W\sbs V$ and $y,W\sat\phi$, since $y,U\sat\Box\phi$, and
we are done, or
$W\sbs U$ and $W\not\sbs V$ so we have
$V\sbs W\sbs U$, since the subsets containing $y$ are linearly ordered. In this
case,  we have $ x\in W$, since $x\in V$. By our assumption
$x,U\sat\Box\sk\phi$, we have $x,W\sat\sk\phi$. So
$y,W\sat\phi$. Now,
$y \in U$ and $y, U\sat \Box \phi $ imply together $y,U\sat \dam\psi \land
\Box\phi$.
\end{prf}

Note that Axiom~\ref{ax:boxk} follows from Axiom~\ref{ax:ntree} (substitute
$\phi$ with $\top$). Axiom~\ref{ax:boxk} has a particular interest; if we
replace $\sk$ with the universal quantifier it becomes the well-known Barcan
formula
$$\forall x \Box\phi(x)\to \Box \forall x \phi(x).$$
Our system (and therefore  $\bf MP$ and $\bf MP^*$, since this formula belongs
to their axiomatization) can be thought as a propositional analogue of a first
order modal system interpreted over varying {\em restricting} domains
(see~\cite{FIT93}).


\subsection{Completeness}

Our proof of completeness is based on a construction of a treelike  model which is
(strongly) equivalent to each generated canonical submodel of the canonical
model of $\bf MPT$.

The {\em canonical model} of $\bf MPT$ is the structure
$$\clc=\left(S,\{\todam,\tol\},v\right),$$
where
$$\begin{array}{clc}
& S=\{s\sbs\cl|s \hbox{ is $\bf MPT$-maximal consistent} \},& \\
& s\todam t \hbox{ iff }\{\phi\in\cl|\Box\phi\in s\}\sbs t, & \\
& s\tol t \hbox{ iff }\{\phi\in\cl|\sk\phi\in s\}\sbs t, & \\
& v(A)=\{s\in S | A\in s\}, &
\end{array}$$
along with the usual satisfaction relation (defined inductively):
$$\begin{array}{lll}
s\csat A & \hbox{iff} & s\in v(A)  \\
s\csat\neg\phi & \hbox{iff} & s\notcsat\phi  \\
s\csat\phi\land\psi & \hbox{iff} & s\csat\phi\hbox{ and }s\csat\psi \\
s\csat\Box\phi & \hbox{iff} & \hbox{for all }t\in S,\
                       s\todam t \hbox{ implies }t\csat\phi \\
s\csat\sk\phi & \hbox{iff} & \hbox{for all }t\in S,\
                       s\tol t \hbox{ implies }t\csat\phi.
\end{array}$$
We write $\clc\sat\phi$, if $s\csat\phi$ for all $s\in S$.

A canonical model exists for
all consistent bimodal systems with the normal axiom scheme for each
modality (as $\bf MPT$).
We have the following well known theorems (see~\cite{CH}, or~\cite{GL}).

\begin{theorem}[Truth Theorem]
For all $s\in S$ and $\phi\in\cl$,
$$s\csat\phi\qquad\hbox{iff}\qquad\phi\in s.$$
\end{theorem}

\begin{theorem}[Completeness Theorem]
For all $\phi\in\cl$,
$$\clc\sat\phi\qquad\hbox{iff}\qquad\vdash_{\bf MPT}\phi.$$
\end{theorem}

We shall now prove some properties of $\clc$.

\begin{prop}
\label{prop:propcan}
\renewcommand{\theenumi}{\alph{enumi}}
\begin{enumerate}
\item \label{s43}
The canonical frame  is reflexive, transitive and  connected with respect
to the relation $\todam$.
\item \label{tolequiv}
The relation $\tol$ is an equivalence relation.
\item
\label{lem:cross}
For all $s,s',t\in S$, if $s\todam s' \tol t$ then there exists $t'\in S$ such
that $s\tol t' \todam t$.
\item
\label{lem:boxandk}
For all $s,s'\in S$, if $s\tol s'$ and $s\todam s'$ then $s=s'$.
\item \label{antisym}
The relation $\todam$ is antisymmetric.
\end{enumerate}
\end{prop}

\begin{prf}
For Part~\ref{s43}, Axioms~\ref{ax:normb} through \ref{ax:s4} and
Axiom~\ref{ax:tree}  characterize reflexive, transitive and connected
frames (these axioms comprise the system $\bf S4.3$).

For Part~\ref{tolequiv}, $\sk$ is axiomatized with the $\bf S5$ axioms.

Part~\ref{lem:cross} is an immediate consequence of Axiom~\ref{ax:boxk}.

To show Part~\ref{lem:boxandk}, let
$$
I_{s,s'} \quad = \quad \{ t \mid s \todam t \todam s' \},
$$
for all pairs $(s,s')$ such that $s\tol s'$ and $s\todam s'$.

We shall prove  by induction on the complexity of $\phi$ that, for all pairs
$(s,s')$ such that $s\tol s'$ and $s\todam s'$,  $\phi$ belongs to some $t\in
I_{s,s'}$ if and only if $\phi$ belongs to s. This shows that $\bigcup
I_{s,s'}\sbs s$. Further,  we have $I_{s,s'}=\{s\}$, since $s\in I_{s,s'}$.
Therefore $s=s'$.

 If $\phi$ is an atomic formula $A$ and $A\in t$, for some $t\in I_{s,s'}$,
then $\dam A\in s$. Therefore,  by axiom~\ref{ax:atom}, $\Box A\in s$. Hence,
$A\in s$.

The cases of negation and conjunction are straightforward.

 If $\phi=\Box\psi$, let $\Box\psi\in t$, for some $t\in I_{s,s'}$.
In particular, $\psi\in t$ and by induction hypothesis, $\psi\in s$.
Suppose, towards a contradiction, that $\dam\neg\psi\in s$. Then there exists
$r\in S$ such that $s\todam r$ and $\neg\psi\in r$. Since the frame is connected,
$s\todam s'$ and $s\todam r$ imply that either $s'\todam r$ or $r \todam s'$.
If $s' \todam r$ then $t \todam r$ which is contradiction, since $\Box\psi\in
t$ and $\neg\psi\in r$. If $r \todam s'$ then, by induction hypothesis,
$\neg\psi\in s$ which is a contradiction, since $\psi\in s$ and $s$ is consistent.
Hence $\Box\psi\in s$.

 If $\phi=\sk\psi$,  let $\sk\psi\in t$ for some $t\in I_{s,s'}$.
Suppose, towards a contradiction, that $\sfl\neg\psi\in s$. Then there exists
$r\in S$ such that $s\tol r$ and
$\neg\psi\in r$.  We have
$s'\tol r$, since $s' \tol s$. Since $t\todam s'$,  there  exists, by
Part~\ref{lem:cross}, $r'\in S$ such that $t \tol r' \todam r$. We have $\psi\in
r'$, since $\sk\psi\in t$. Since
$s\todam t \tol r'$,  there  exists, by Part~\ref{lem:cross}, $r''\in S$ such
that
$s\tol r'' \todam r'$. Notice that $r'' \todam r' \todam r$ and $r''\tol r$,
and so $r'',r',r\in I_{r'',r}$. By our previous assumption, we have
$\neg\psi\in r$ and $\psi\in r'$. By induction hypothesis on $I_{r'',r}$, both
$\neg\psi$ and $\psi$ should belong to $r''$ which is a contradiction to its
consistency.

For Part~\ref{antisym}, we shall prove by induction on the structure of $\phi$
that, for all $s,t\in S$ such that $s\todam t\todam s$,  $\phi\in s$ if and only
$\phi\in t$.

The cases of atomic formula, negation, conjunction and $\Box$ are
straightforward. We shall show the $\phi=\sk\psi$ step. Let $\sk\psi \in s$, and
suppose $\sfl\neg\psi \in t$ towards a  contradiction. Then there exists $r\in
S$ such that $t\tol r$ and $\neg\psi\in r$. Since $s\todam t\tol r$, there exists
$p\in S$ such that $s\tol p \todam r$. Also, $\psi \in p$, since $\sk\psi \in S$.
Now, since $t\todam s \tol p$ there exists $r'\in S$ such that $t\tol r' \todam
p$. This implies $r'\todam p \todam r$ and $r\tol r'$. Therefore,
by Part~\ref{lem:boxandk}, $r=r'$. Thus we have $r\todam p \todam r$ with
$\neg\psi\in r$ and $\psi\in p$ which is a contradiction to the induction
hypothesis.
\end{prf}

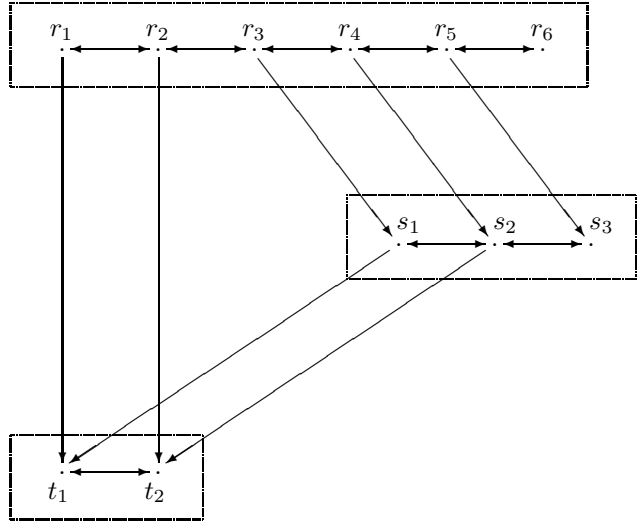
\begin{figure}[t]
\centering
\setlength{\unitlength}{0.00083300in}%
\begin{picture}(3924,3249)(889,-2998)
\thinlines
\put(1201,-61){$.$}
\put(1801,-61){$.$}
\put(2401,-61){$.$}
\put(3001,-61){$.$}
\put(3601,-61){$.$}
\put(4201,-61){$.$}
\put(1276,-51){\vector(-1, 0){  0}}
\put(1276,-51){\vector( 1, 0){490}}
\put(1876,-51){\vector(-1, 0){  0}}
\put(1876,-51){\vector( 1, 0){490}}
\put(2476,-51){\vector(-1, 0){  0}}
\put(2476,-51){\vector( 1, 0){490}}
\put(3076,-51){\vector(-1, 0){  0}}
\put(3076,-51){\vector( 1, 0){490}}
\put(3676,-51){\vector(-1, 0){  0}}
\put(3676,-51){\vector( 1, 0){490}}
\put(2446,-106){\vector( 3,-4){834}}
\put(3046,-106){\vector( 3,-4){834}}
\put(3646,-106){\vector( 3,-4){834}}
\put(3301,-1276){$.$}
\put(3901,-1276){$.$}
\put(4501,-1276){$.$}
\put(3266,-1306){\vector(-3,-2){1995}}
\put(3866,-1306){\vector(-3,-2){1995}}
\put(1226,-106){\vector( 0,-1){2525}}
\put(1826,-106){\vector( 0,-1){2525}}
\put(1276,-2691){\vector(-1, 0){  0}}
\put(1276,-2691){\vector( 1, 0){490}}
\put(3376,-1266){\vector(-1, 0){  0}}
\put(3376,-1266){\vector( 1, 0){490}}
\put(3976,-1266){\vector(-1, 0){  0}}
\put(3976,-1266){\vector( 1, 0){490}}
\put(1201,-2701){$.$}
\put(1801,-2701){$.$}
\put(901,-286){\dashbox{10}(3600,525){}}
\put(3001,-1486){\dashbox{10}(1800,525){}}
\put(901,-2986){\dashbox{10}(1200,525){}}
\put(3311,-1171){$s_1$}
\put(3911,-1171){$s_2$}
\put(4511,-1171){$s_3$}
\put(1136,-2856){$t_1$}
\put(1736,-2856){$t_2$}
\put(1146, 31){$r_1$}
\put(1746, 31){$r_2$}
\put(2346, 31){$r_3$}
\put(2946, 31){$r_4$}
\put(3546, 31){$r_5$}
\put(4146, 31){$r_6$}
\end{picture}
\caption{A generated submodel of the canonical model.}
\label{fig:canonicalexample1}
\end{figure}

The canonical model is not a (model corresponding to a) treelike model. A
counterexample will appear later on (see Figure~\ref{fig:canonicalexample1}).
However, by defining a number of equivalence relations, we shall be able to
construct a treelike model equivalent to each generated part of the canonical
model.

For all $t\in S$, let
$[t]=\{s\in S\mid s\tol t\}$, i.e. the equivalence class under $\tol$ where $t$
belongs. Let $\clc_\fsk=\{[t]\mid t\in S\}$.
We define the following relation on $\clc_\fsk$.
$$
[t_1] \leq [t_2]\quad \hbox{iff}\quad \hbox{there exist}\ s_1,s_2\in S \
\hbox{such that}\ s_1\in[t_1],s_2\in[t_2]\ \hbox{and}\ s_2\todam s_1.
$$

\begin{prop}
The relation $\leq$ is a partial order.
\end{prop}

\begin{prf}
Since $t\todam t$, we have $[t]\leq [t]$ and reflexivity follows.

For antisymmetry, let $[t_1] \leq [t_2]$ and $[t_2] \leq [t_1]$ for some
$t_1,t_2\in S$. Then there exist $s_1, s_2, s'_1,s'_2 \in S$ such that
$s_1,s'_1\in [t_1]$, $s_2, s'_2\in [t_2]$, $s_2\todam s_1$ and $s'_1\todam
s'_2$. Since $s_2 \todam s_1 \tol s'_1$, there exists $s''_2\in S$ such that
$s_2\tol s''_2 \todam s'_1$. So we have $s''_2 \todam s'_1 \todam s'_2$ and
$s''_2\tol s'_2$ which implies,
by~Proposition~\ref{prop:propcan}(\ref{lem:boxandk}),
$s''_2=s'_2$. Therefore $s'_1=s'_2$, by $\todam$'s antisymmetry.
Hence $[t_1]=[s'_1]=[s'_2]=[t_2]$.

For transitivity, let $[t_3] \leq [t_2] \leq [t_1]$ for some $t_1,t_2,t_3\in S$.
Then there exist $s_1\in[t_1]$, $s_2,s'_2\in[t_2]$, and $s_3\in [t_3]$ such that
$s_1\todam s_2$ and $s'_2\todam s_3$. Since $s_1\todam s_2 \tol s'_2$, there
exists $s'_1\in S$ such that $s_1 \tol s'_1 \todam s'_2$. So $s'_1\todam s_3$,
and therefore $[t_3]=[s_3]\leq [s'_1]=[t_1]$.

\end{prf}

 A subset $X$ of $S$, the domain of the canonical model $\clc$, is
called {\em $\sk\Box$-closed} whenever
$$\hbox{ if } s\in X, \hbox{ and } s\todam t \hbox{ or } s\tol t,
\quad\hbox{then}\quad t\in X.$$
The intersection of $\sk\Box$-closed sets is still $\sk\Box$-closed, therefore
we can define the smallest $\sk\Box$-closed containing $t$, for all
$t\in S$. We shall denote this set by $S^t$.
Fix ${t_0}\in S$. We define the model
$$\clc^{t_0}=\left(
S^{t_0},\todam|_{S^{t_0}\times S^{t_0}},\tol|_{S^{t_0}\times S^{t_0}},v^{t_0}
\right),$$ where $\todam|_{S^{t_0}\times S^{t_0}}$, $\tol|_{S^{t_0}\times
S^{t_0}}$ and
$v^{t_0}$ are the restrictions of $\todam$, $\tol$ and $v$ to $S^{t_0}\times
S^{t_0}$ and $S^{t_0}$ respectively. We shall call this model the {\em submodel
of
$\clc$ generated by
$t_0$}.

Observe that if we restrict the partial order $\leq$ to $\clc^{t_0}$ then
$[t_0]$ is the greatest element under $\leq$.

For each generated submodel of the canonical model, we shall construct a treelike
model which is equivalent to it.

For each $s\in S^{t_0}$, let
$$
\lmean s \rmean\quad=\quad \{t\in [t_0] \mid \hbox{there exists} \ t'\in [s]\
\hbox{such that}\ t\todam t'\}.
$$
Notice that $\lmean s \rmean\sbs [t_0]$.

For each $s\in S^{t_0}$, we define the following  relation $\sim_s$ on
$\lmean s \rmean$
$$
t_1\sim_s t_2 \quad \hbox{iff} \quad \hbox{for all}\ [s]\leq[s'],\ t_1\in\lmean
s' \rmean\ \hbox{iff}\ t_2\in\lmean s' \rmean .$$

\begin{prop}
For all $s\in S^{t_0}$, the relation $\sim_s$ is an equivalence relation.
\end{prop}

\begin{prf}
This is because $\sim_s$ inherits the properties of $\tol$.
\end{prf}

We denote the equivalence class of $t$ under $\sim_s$ with $[t]_s$.
We have $[t]_s\sbs \lmean s \rmean\sbs [t_0]$.

Let $\langle X,\co^{t_0} \rangle$ be the subset space where
$$
X\quad=\quad\{t\mid t\in [{t_0}] \}
$$
and
$$
\co^{t_0}\quad=\quad \{[t]_s\mid t\in \lmean s \rmean\ \hbox{and}\ s\in S^{t_0} \}.
$$
It is clear that $\co^{t_0}\sbs \cp(X)$.

\begin{lemma}\label{lem:eqsbs}
If $[s_1]\leq [s_2]$ and
$t\in \lmean s_1 \rmean \cap \lmean s_2 \rmean $ then $[t]_{s_1}\sbs [t]_{s_2}$.
\end{lemma}

\begin{prf}
Immediate from  the definition of $\sim_s$ .
\end{prf}

To elaborate the above process, we present the following simple example.

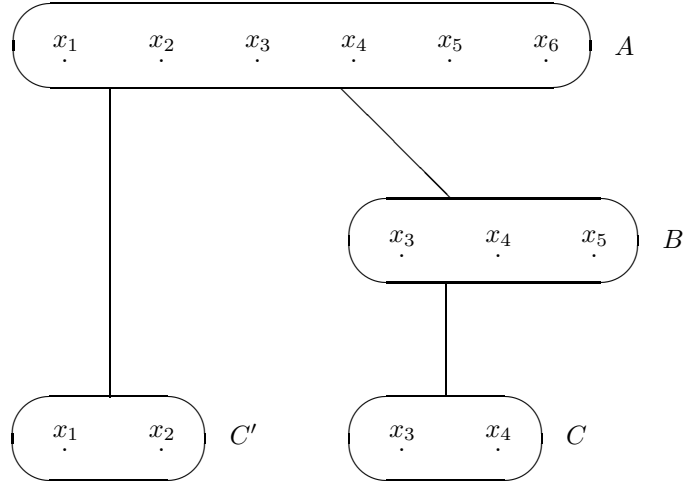
\begin{figure}[t]
\centering
\setlength{\unitlength}{0.00083300in}%
\begin{picture}(4062,3250)(889,-2999)
\thinlines
\put(1201,-61){$.$}
\put(1801,-61){$.$}
\put(2401,-61){$.$}
\put(3001,-61){$.$}
\put(3601,-61){$.$}
\put(4201,-61){$.$}
\put(1501,-226){\line( 0,-1){1928}}
\put(2946,-228){\line(1,-1){690}}
\put(3601,-1441){\line( 0,-1){710}}
\put(3301,-1276){$.$}
\put(3901,-1276){$.$}
\put(4501,-1276){$.$}
\put(2701,41){\oval(3600,525)}
\put(3896,-1176){\oval(1800,525)}
\put(1201,-2491){$.$}
\put(1801,-2491){$.$}
\put(1496,-2411){\oval(1200,525)}
\put(3301,-2491){$.$}
\put(3901,-2491){$.$}
\put(3596,-2411){\oval(1200,525)}
\put(1146, 31){$x_1$}
\put(1746, 31){$x_2$}
\put(2346, 31){$x_3$}
\put(2946, 31){$x_4$}
\put(3546, 31){$x_5$}
\put(4146, 31){$x_6$}
\put(3246,-1186){$x_3$}
\put(3846,-1186){$x_4$}
\put(4446,-1186){$x_5$}
\put(1146,-2401){$x_1$}
\put(1746,-2401){$x_2$}
\put(3246,-2401){$x_3$}
\put(3846,-2401){$x_4$}
\put(4651,-16){$A$}
\put(4951,-1221){$B$}
\put(4351,-2446){$C$}
\put(2251,-2446){$C'$}
\end{picture}
\caption{The treelike model corresponding to  Figure~\ref{fig:canonicalexample1}}
\label{fig:canonicalexample2}
\end{figure}

\noindent {\em Example:\/}
A part of the canonical model appears in Figure~\ref{fig:canonicalexample1}.
(Horizontal and downward arrows correspond to $\tol$ and $\todam$, respectively.)
We would like to make subsets of a treelike space correspond to equivalence
classes under $\tol$. Canonical model worlds related with $\todam$ will be
represented by a single point. However, this model is not a treelike model:
$\{r_1,t_1\}$ and $\{r_3,s_1,t_1\}$ should make two distinct points. To remedy
that, we ``trace back'' each equivalence class under $\tol$ to the uppermost one.
For instance, $[t_1]=\{t_1,t_2\}$ is traced back to $[r_1]=\{r_1,r_2,r_3,r_4\}$.
The latter forms $\lmean t_1 \rmean$. Next, we split $\lmean t_1 \rmean$ into
equivalence classes under $\sim_{t_1}$, i.e. $[r_1]_{t_1}=\{r_1,r_2\}$ and
$[r_3]_{t_1}=\{r_3,r_4\}$, since $r_1 \sim_{t_1} r_2$ and $r_3 \sim_{t_1} r_4$.
Finally, we replace $[t_1]$ with as many copies as these equivalence classes (see
Figure~\ref{fig:canonicalexample2}).
The infinite case is taken care of by
Lemma~\ref{lem:maxlo}. The resulting space (of
Figure~\ref{fig:canonicalexample2}) is a treelike space. Note that we could
have replaced this procedure by one that employs maximal branches but we find the
present one simpler.

\begin{prop}
\label{prop:treespace}
The subset space $\langle X,\co^{t_0} \rangle$ is a treelike space.
\end{prop}

\begin{prf}
Suppose $[t_1]_{s_1}\cap [t_2]_{s_2}\not= \emptyset$. Let
$t\in [t_1]_{s_1}\cap [t_2]_{s_2}$. We have either $s_1 \todam s_2$ or
$ s_2 \todam s_1$, since $t \todam s'_1$, $t\todam s'_2$,
for some $s'_1\in [s_1]$ and $s'_2\in [s_2]$,  and
the canonical frame is connected. The former implies $[s_1] \leq [s_2]$. Thus, by
Lemma~\ref{lem:eqsbs}, $ [t_1]_{s_1} = [t]_{s_1} \sbs [t]_{s_2} =
[t_2]_{s_2}$. Similarly, the latter implies $ [t]_{s_2} \sbs [t]_{s_1} $.
\end{prf}

Let $\langle X,\co^{t_0}, i \rangle$ be the treelike  model where $X$ and
$\co^{t_0}$ are as above, and $i(A)=v^{t_0}(A)$ where $v^{t_0}$ is the initial
interpretation restricted on
$\clc^{t_0}$.

An element of $X \dot{\times} \co^{t_0}$ can have more than one representation.
In order  to prove the semantical equivalence we are opting for, we shall
choose a canonical representation. So, given a pair
$( t, [t']_{s'}) \in  X\dot{\times} \co^{t_0}$, its {\em canonical
representation}
is  $( t , [t]_s)$ where  $s$ is such that $ t \todam s \tol s'$. Its existence
is assured by the definition of $[t']_{s'}$ and uniqueness by
Proposition~\ref{prop:propcan}(\ref{lem:boxandk}).
From now on, we shall use the canonical representation wherever is possible.

\begin{lemma}\label{lem:maxlo}
Let $t\in [t_0]$ and $s\in S^{t_0}$ such that $t \todam s$. Then for all $s'
\in [s]$ there exists $t'\in [t_0]$ such that $t' \todam s'$ and $t\sim_s t'$,
i.e. $t' \in [t]_s$.
\end{lemma}

\begin{prf}
Let
$$\{ t_i \}_{i\in I} $$ be the linear order of all members of $S^{t_0}$
under $\todam$ such that $t \todam t_i \todam s $.

Now, let
$$
\begin{array}{llll}
 T' & = &  & \{ \dam\psi \mid \psi \in s' \} \\
    &   &  \cup & \{\chi \mid \sk\chi\in t \} \\
    &   &  \cup & \{\dam\go \mid \sk\go\in t_i,\ \hbox{for some}\ i\in I \} \\
    &   &  \cup & \{\Box\phi \mid \Box\sk\phi\in t \ \hbox{and}\ \Box\phi\in
    s' \}.
\end{array}
$$
$T'$ is consistent. For if not, then there would be
$\psi, \go_{1}, \go_{2},\ldots,\go_{n},\chi,\phi$ as above with
$i_1 , i_2, \ldots, i_n \in I$ and $i_1 \leq i_2 \leq \ldots \leq i_n$ such that
$$
\MPT \dam\psi\land\bland_{k=1}^{n}\dam\go_{k}\land\chi\to\dam\neg\phi.
$$
Thus
$$
\MPT
\sk\left(
\dam\psi\land\bland_{k=1}^{n}\dam\go_{k}\land\chi\to\dam\neg\phi\right).
$$
We shall prove that the negation of the above formula belongs to $t$
and reach a contradiction.
Since
$\psi\land\Box\phi\in s'$, we have $\sfl(\psi\land\Box\phi)\in s$.
Hence
$$\dam\sfl(\psi\land\Box\phi)\in t_{i_1}.$$
Observe that $\Box\sk\phi\in t_{i_1}$ so, by applying axiom~\ref{ax:ntree},
we have  $$\sfl(\dam\psi\land\Box\phi)\in t_{i_1}.$$
Since $\sk\go_{1}\in t_{i_1}$, we have
$$\sfl\left(\dam\psi\land\go_{1}\land\Box\phi\right)\in t_{i_1}.$$
Also, $\Box\sk \phi \in t_{i_2}$ and
$$\dam\sfl\left(\dam\psi\land\go_{1}\land\Box\phi\right)\in t_{i_2}.$$
So, by axiom~\ref{ax:ntree},
$$\sfl\left( \dam(\dam\psi\land\go_{1})\land\Box\phi\right)\in t_{i_2}.$$
Since $\sk\go_{2}\in t_{i_2}$, we have
$$\sfl\left(\dam\psi\land\go_{2}\land\dam\go_{1}\land\Box\phi\right)\in
t_{i_2}.$$
Also, $\Box\sk \phi \in t_{i_3}$ and
$$\dam\sfl\left(\dam\psi\land\go_{2}\land\dam\go_{1}\land\Box\phi\right)\in
t_{i_3}.$$
So, by axiom~\ref{ax:ntree},
$$\sfl\left(\dam
(\dam\psi\land\go_{2}\land\dam\go_{1})\land\Box\phi\right)\in t_{i_3},$$
i.e.
$$\sfl\left(\dam\psi\land\dam\go_{2}\land\dam\go_{1}\land\Box\phi\right)\in
t_{i_3}.$$
Arguing this way and by repeated applications of
axiom~\ref{ax:ntree} we have
$$\sfl\left(
\dam\psi\land\bland_{k=1}^{n}\dam\go_{k}\land\Box\phi\right)\in t.$$
Since $\sk\chi\in t$, we have
$$\sfl\left(
\dam\psi\land\bland_{k=1}^{n}\dam\go_{k}\land\chi\land\Box\phi\right)\in t$$
which is the negation of the formula that $\bf MPT$ proves. Therefore $T'$ is
consistent. Let $t'$ be a maximal extension of $T'$.

We shall show that $t'$ is the required theory of the lemma.
We begin by showing that if $t'\todam r' \todam s'$
then $r' \tol t_i$, for some $i\in I$, i.e. $t\in \lmean r' \rmean$.
So suppose that $t'\todam r' \todam s'$. If $r'=s'$ we are done. If not, let
$$R=\{\psi\mid\Box\psi\in t \}\cup \{\sfl\chi\mid\chi\in r'\}.$$
$R$ is consistent. For if not, then there would be $\psi$ and $\chi$ as above such
that
$$\MPT \psi\to\neg\sfl\chi.$$
Since $r'\todam s'$ and $r'\not= s'$, there exists $\chi' \in r'$ such that
$\Box\neg\chi' \in s'$. Let $\phi=\chi\land\chi'$. Observe that
$\Box\neg(\chi\land\chi')\in s'$, i.e. $\Box\neg\phi\in s'$, and $\phi\in r'$.
Further,
$$\MPT \psi\to\neg\sfl(\chi\land\chi'),$$
i.e.
$$\MPT \psi\to\neg\sfl\phi,$$
and therefore,
$$\MPT \Box\psi\to\Box\sk\neg\phi.$$
Now, we have  $\Box\sk\neg\phi\in t$ and $\Box\neg\phi\in s'$, since $\Box\psi\in
t$. By definition of $T'$ above, we have
$\Box\neg\phi\in T'$, and therefore $\Box\neg\phi\in t'$ ($t'$ is an extension of
$T'$). In this case, $\neg\phi\in r'$ which is a contradiction.
Therefore $R$ is consistent. So a maximal extension $r$ of $R$ has the property
$t\todam r \todam s$. Hence $r=t_i$, for some $i\in I$.

We must now prove that  $t' \in \lmean t_i \rmean$, for all $i\in I$.
Let
$$T'_i=\{\psi \mid \Box\psi\in t' \} \cup \{\omega \mid \sk\omega \in t_i\}.$$
$T'_i$ is consistent. If not, then
$$\MPT \psi\to \neg\omega,$$
for some $\phi$ and $\omega$ as above,
which implies
$$\MPT \Box\psi\to \Box\neg\omega,$$
i.e.
$$\MPT \Box\psi\to \neg\dam\omega.$$
So $\neg\dam\omega\in t'$, since $\Box\psi\in t'$. But, by  definition,
$\dam\omega\in T'\sbs t'$ which is a contradiction. Therefore a maximal extension
$t'_i$ of $T'_i$ is such that $t'\todam t'_i \tol t_i$. Hence $t'\in \lmean t_i
\rmean$.

Combining the above proofs we have $t \sim_s t'$.

\end{prf}

We now have the following theorem.

\begin{theorem}
\label{thm:iso}
For all $s\in S^{t_0}$ and $t\in X$ such that $t \todam s$,
$$
\phi \in s \quad \hbox{iff} \quad t,[t]_s \sat \phi.
$$
\end{theorem}

\begin{prf}
By induction on the structure of $\phi$.
For an atomic formula $A$, we have  that  $t\in i(A)$ if and only if
$s\in i(A)=v^{t_0}(A)$, i.e. $A\in s$, because of Axiom~\ref{ax:atom} and $t
\todam s$.

Negation and conjunction are straightforward.

Suppose $\phi=\Box\psi$. Let $\Box\psi\in s$ and $t,[t]_s\sat \dam\neg\psi$,
for some $s$ and $t$ as in the theorem's statement. This implies that there
exists $s'\in S^{t_0}$ such that $[t]_{s'}\sbs [t]_s$, $t \todam s'$ and
$t,[t]_{s'}\sat\neg\psi$. By induction hypothesis, $\neg\psi\in s'$. We have now
that $t\todam s$ and $t \todam s'$ which, by connectivity, implies either
$s'\todam s$ or $s\todam s'$. In the former case, we have $[s]\leq [s']$ and
hence, by Lemma~\ref{lem:eqsbs}, $[t]_s \sbs [t]_{s'}$. So $[t]_s = [t]_{s'}$.
Therefore $s=s'$, by~Proposition~\ref{prop:propcan}(\ref{lem:boxandk}), which
is a contradiction to our
 hypothesis ($\phi\in s$). In the latter case, we
have $s\todam s'$ which again contradicts our hypothesis  ($\phi\in s$).

For the other direction, suppose that $t,[t]_s\sat \Box\psi$ and
$\dam\neg\psi \in s $ for some  $s$ and $t$ as above. Then there exists
$s'\in S^{t_0}$ such that $s\todam s'$ and $\neg\psi\in s'$. Thus,
$t,[t]_{s'}\sat\neg\psi$ by induction hypothesis. Moreover $[t]_{s'}\sbs [t]_s$
by Lemma~\ref{lem:eqsbs}, which is a contradiction.

If $\phi=\sk\psi$, let $\sk\psi\in s$ and suppose $t,[t]_s\sat \sfl\neg\psi$,
for some $s$ and $t$ as in the theorem's statement, towards a contradiction.
Then there exists $t'\in[t]_s$ such that $t',[t]_s\sat\neg\psi$, i.e.
$t',[t']_{s'}\sat\neg\psi$, for some $s'\in S^{t_0}$ such that $t'\todam s'$
and  $s'\tol s$, which is a contradiction.

For the other direction, suppose that $t,[t]_s\sat \sk\psi$ and
$\sfl\neg\psi \in s $, for some  $s$ and $t$ as above. Then there exist
$s'\in S^{t_0}$ such that $s\tol s'$ and $\neg\psi\in s'$. By
Lemma~\ref{lem:maxlo}, there exists $t'\in [t]_s$ such that $t'\todam s'$. Then
we have $t',[t]_{s}\sat\neg\psi$ by induction hypothesis. Therefore $t,[t]_s\sat
\neg\sk\psi$ which is a contradiction.

\end{prf}

Combining now Proposition~\ref{prop:treespace} and Theorem~\ref{thm:iso} we have
the following

\begin{corol}
The system $\bf MPT$ is complete with respect to treelike spaces.
\end{corol}

\subsection{Decidability}

For each treelike  model and formula $\phi$, we shall construct an equivalent finite
subset space
of bounded size with respect to the complexity of $\phi$. This is a
kind of ``semantic'' filtration, based on  geometric properties of
treelike models, using a technique first introduced in~\cite{KG1}.

In the following we assume that $\langle X,\co
\/\rangle$ is a treelike space.
Our aim  is to find a partition of $\co$, where a given
formula $\phi$ ``retains its truth value'' for each point throughout a
member of this partition. It turns out that there exists a finite
 partition of this kind.

First we need some definitions. (Note that the following hold, although we refer to
a treelike space $\co$, for  an
arbitrary family of subsets of $X$.)

\begin{dfn}
Given a finite family $\cf=\{ U_1,\ldots,U_n\}\sbs\cp(X)$, i.e. of  subsets of
$X$, we define the {\em remainder} of (the principal ideal in $(\co,\sbs)$
generated by) $U_k$ by
$$\Rem{\cf}{U_k}\quad=\quad\rem{U_k}{U_k\not\subseteq U_i}{}{U_i},$$
where $\dar U_k=\{V\in\co\mid V\sbs U_k\}$. Note that $\Rem{\cf}{U_k}\sbs\co$
(but not necessarily $U_k\in \co$).
\end{dfn}

\begin{prop} In a finite family $\cf=\{U_1,\ldots,U_n\}\sbs\cp(X)$ closed
under intersection, we have
$$\Rem{\cf}{U_i}\quad=\quad\rem{U_i}{U_j\subset U_i}{}{U_j},$$ for
$i=1,\ldots,n$. \end{prop}

\begin{prf}
$$
\begin{array}{rcl}
\Rem{\cf}{U_i} & = & \dar U_i - \bigcup_{U_i\not\subseteq U_h}\dar U_h \\
             & = & \dar U_i - \bigcup_{U_i\not\subseteq U_h}\dar (U_h\cap U_i)\\
             & = & \dar U_i - \bigcup_{U_j\subset U_i}\dar U_i. \\
\end{array}
$$
\end{prf}

We denote $\bigcup_{U_i\in\cf}\dar U_i$ with $\dar\cf$.

\begin{prop} If $\cf=\{U_1,\ldots,U_n\}$ is a finite family of subsets of $X$
closed under intersection then
\label{prop:part}
\renewcommand{\theenumi}{\alph{enumi}}
\begin{enumerate}
\item \label{part1}
$\Rem{\cf}{U_i}\cap\Rem{\cf}{U_j}=\emptyset$, for $i\not= j$,
\item \label{part2}
$\bigcup^n_{i=1}\Rem{\cf}{U_i}=\dar\cf$,
i.e.
$\{\Rem{\cf}{U_i}\}^n_{i=1}$ is a partition of $\dar\cf$. From now on we shall
call a finite family of subsets
$\cf$ closed under intersection a {\em finite partition (of $\dar\cf$),}
\item
if $V_1,V_2\in\co$, $V_1\in\Rem{\cf}{U_i}$ and $V_1\sbs V_2\sbs U_i$
then $V_2\in\Rem{\cf}{U_i}$, i.e. $\Rem{\cf}{U_i}$ is convex,\label{convex}
\item
if $\{V_j\}_{j\in J}\sbs \Rem{\cf}{U_i}$ then $\bigcup_{j\in J} U_j \sbs U_i$.
\label{union}
\end{enumerate}
\end{prop}

\begin{prf}
Parts~\ref{part1}, \ref{convex} and \ref{union} are immediate from the definition.

For Part~\ref{part2}, suppose that $V\in\dar\cf$ then
$V\in\Rem{\cf}{\bigcap_{V\in\dar U_i}U_i}$.
\end{prf}

Every partition of a set induces an equivalence relation on this set.
The members of the partition comprise the equivalence classes. We
denote the equivalence relation induced by $\cf$ by $\sim_\cf$.

\begin{dfn}
Given a set of  subsets $\cg$, we define the relation $\sim'_\cg$
on $\co$ with $V_1\sim'_\cg V_2$ if and only if $V_1\sbs U\eqv V_2\sbs
U$ for all $U\in\cg$.
\end{dfn}

We have the following

\begin{prop}
The relation $\sim'_\cg$ is an equivalence.
\end{prop}

\begin{prop}
Given a finite partition $\cf$, we have $\sim'_\cf = \sim_\cf$ i.e. the
remainders of $\cf$ are the equivalence classes of $\sim'_\cf$.
\label{prop:equiv}\end{prop}

\begin{prf}
Suppose $V_1\sim'_\cf V_2$ then $V_1$ and $V_2$ belong  to
$\Rem{\cf}{U}$ where
$$U=\bigcap\{ U' | V_1,V_2\sbs U,\ U'\in\cf\}.$$
For the opposite direction, suppose $V_1,V_2\in\Rem{\cf}{U}$ and there
exists $U'\in\cf$ such that $V_1\sbs U'$ while $V_2\not\sbs U'$.
Then we have $V_1\sbs U'\cap U$, $U'\cap U\in\cf$ and
$U'\cap U\subseteq U$ i.e. $V_1\not\in\Rem{\cf}{U}$.
\end{prf}

\begin{prop}
If $\cg$ is a finite set of subsets of $X$ then ${\sf Cl(\cg)}$,
its closure under intersection, is a finite partition for $\dar\cg$.
\label{prop:closinter}
\end{prop}

The last proposition enables us to give yet another characterization
of remainders: every family of points in a complete lattice
closed under arbitrary joins comprises a
{\em closure system}, i.e. a set of fixed points of a closure operator
of the lattice (cf. \cite{COMP}.) Here the lattice is the powerset of $X$. If
we restrict ourselves to a finite
number of fixed points then we just ask for a finite set of subsets closed under
intersection i.e. Proposition~\ref{prop:closinter}.
Thus a closure operator in the lattice of the powerset of $X$ induces an
equivalence relation to any family of subsets of $X$.
Two subsets are equivalent if they have the same closure, and the equivalence
classes of this relation are just the remainders of the  subsets
which are fixed points  of the closure operator.

We now introduce the notion of stability corresponding to what we
mean by ``a formula retains its truth value on a set of subsets''.

\begin{dfn}
Let $\cg\sbs\co$  then $\cg$ is {\em stable for
$\phi$}, if for all $x$, either $x,V\sat\phi$ for all $V\in\cg$,
or $x,V\sat\neg\phi$ for all $V\in\cg$.
\end{dfn}

\begin{prop}
Let $\cg_1$,$\cg_2\sbs\co$  then
\label{prop:rem}
\renewcommand{\theenumi}{\alph{enumi}}
\begin{enumerate}
\item if $\cg_1\sbs\cg_2$ and $\cg_2$ is stable for
$\phi$ then $\cg_1$ is stable for $\phi$, and\label{subrem}
\item
if $\cg_1$ is stable for $\phi$ and $\cg$ is stable for $\chi$ then
$\cg_1\cap\cg_2$ is stable for $\phi\land\chi$.\label{interem}
\end{enumerate}
\end{prop}

\begin{prf} Part~\ref{subrem} is easy to see while Part~\ref{interem} is a
corollary of Part~\ref{subrem}.
\end{prf}

\begin{dfn}
A finite partition $\cf=\{U_1,\ldots,U_n\}$ is called
a {\em stable partition for} $\phi$, if $\Rem{\cf}{U_i}$ is stable
for $\phi$, for all $U_i\in\cf$.
\end{dfn}

\begin{prop} If $\cf=\{U_1,\ldots,U_n\}$ is a stable partition for
$\phi$, so is
$${\cf}'={\sf Cl}(\{U_0,U_1,\ldots,U_n\}),$$
where $U_0\in\dar\cf$.\label{prop:unfsp}  \end{prop}

\begin{prf}
Let $V\in\cf'$, then there exists $U_l\in\cf$ such that
$\Rem{\cf'}{V}\sbs\Rem{\cf}{U_l}$ (e.g.
$U_l=\bigcap\{U_i|U_i\in\cf,V\sbs U_i\}$), i.e. $\cf'$ is a
{\em refinement} of $\cf$. But $\Rem{\cf}{U_l}$ is stable for $\phi$
and so is $\Rem{\cf'}{V}$
by Proposition~{\ref{prop:rem}(\ref{subrem})}.
\end{prf}

The above proposition says  that a
finite stable partition for a treelike space $\co$ remains stable if we
``refine'' it.

 The following is the main theorem of this
section. It says that for each formula $\phi$ we can find a stable partition for
$\phi$ which is essentially a refinement of the stable partition corresponding to
the subformulae of $\phi$.

\begin{theorem}[Partition Theorem]
Let $\cm=\langle X,\co,i\rangle$ be a  treelike  model. Then there
exists  a family $\{\cf^\psi\}_{\psi\in\cl}$
of finite stable partitions such that if $\phi$ is a
       subformula of $\psi$ then $\cf^{\phi}\sbs\cf^{\psi}$ and $\cf^{\psi}$
       is a finite stable partition for $\psi$.\label{thm:main}
\end{theorem}

\begin{prf}
By induction on the structure of the formula $\psi$. In
each step we refine the partition of the induction
hypothesis.
For each $U\in\cf^\psi$, let  $U^\psi=\{x\in
U : x,U\sat\psi\}$. This set determines completely the satisfaction of
$\psi$ on $\Rem{\cf^\psi}{U}$ whenever $\cf^\psi$ is stable.

\begin{itemize}

\item If $\psi=A$ is an atomic formula then
$\cf^A=\{X\}=\{i(\top)\}$, since $\co$ is stable for
all atomic formulae.  We have
$X^A=i(A)$.

\item If $\psi=\neg\phi$ then let $\cf^{\psi}=\cf^{\phi}$,
since the statement of the theorem is symmetric with respect to
negation. We also have
$U^\psi=(X-U^\phi)\cap U$, for all $U\in\cf^\psi$.

\item If $\psi=\chi\land\phi$, let
$$\cf^{\psi}={\sf Cl}(\cf^{\chi}\cup\cf^{\phi}).$$
Observe that
$\cf^{\chi}\cup\cf^{\phi}\sbs\cf^{\chi\land\phi}$.
Now, $\cf^{\psi}$ is a stable partition for $\chi\land\phi$ containing $X$,
since it is a refinement of both $\cf^\chi$ and $\cf^\phi$. Thus, $\cf^{\psi}$ is
a finite stable partition for $\psi$ containing $X$.

\item Suppose $\psi=\sk\phi$. Then, by induction
hypothesis, there exists a finite stable
partition $\cf^{\phi}=\{U_1,\ldots,U_n\}$
for $\phi$ containing $X$.

Now, if $V\in\Rem{\cf^{\phi}}{U_i}\cap\dar U^\phi_i$, for some
$i\in\{1.\ldots,n\}$, then $x,V\sat\phi$, for
all $x\in V$, by definition of $U^\phi_i$. Hence $x,V\sat\sk\phi$, for all
$x\in V$.

On the other hand, if $V\in\Rem{\cf^{\phi}}{U_i}-\dar U^\phi_i$ then there
exists $x\in V$ such that $x,V\sat\neg\phi$ (otherwise $V\sbs U^\phi_i$).
Thus we have $x,V\sat\neg\sk\phi$, for all $x\in V$.
Hence $\Rem{\cf^{\phi}}{U_i}\cap\dar U^\phi_i$ and
$\Rem{\cf^{\phi}}{U_i}-\dar U^\phi_i$ are stable for $\sk\phi$.
Thus the set
$$
\begin{array}{lll}
F & = & \{\Rem{\cf}{U_i}|\ U^\phi_i\not\in\Rem{\cf}{U_i}\}\cup \\
  &   & \{\Rem{\cf}{U_j}-\dar U^\phi_j,\Rem{\cf}{U_j}\cap\dar U^\phi_j|\
U^\phi_j\in U_j\}
\end{array}
$$
is a partition of $\co$ and its members are stable for $\sk\phi$.
Let
$$
\cf^{\fsk\phi}={\sf Cl}(\cf^{\phi}\cup U^\phi_i).$$
We have that $\cf^{\fsk\phi}$ is a finite set of opens and
$\cf^{\phi}\sbs\cf^{\fsk\phi}$. Thus $\cf^{\fsk\phi}$ is  finite
and contains $X$. We have only to prove that $\cf^{\fsk\phi}$
is a stable partition for $\sk\phi$, i.e. every remainder of an open
in $\cf^{\fsk\phi}$ is stable for $\sk\phi$. But for that, observe that
$\cf^{\fsk\phi}$ is a refinement of $F$.
Therefore  $\cf^{\fsk\phi}$ is a finite stable partition
for $\sk\phi$, using Proposition~{\ref{prop:rem}(\ref{subrem})}.

Now, if $U\in\cf^\psi$ then
either $U^{\fsk\phi}=U$ or $U^{\fsk\phi}=\emptyset$.

\item Suppose $\psi=\dam\phi$. Then, let
$$\cf^{\dam\phi}=\cf^{\phi},$$
where $\cf^\phi$ is a finite stable partition for $\phi$ by induction
hypothesis.

We shall show that $\cf^\phi$ is also a finite stable spitting for
$\dam\phi$. Pick $U\in\cf^\phi$ and $x\in U$. If $x,V\sat\neg\phi$, for all
$V\sbs U$ such that $x\in V$, we are done, since
$x,V\sat\neg\dam\phi$. If
 $x,V\sat\phi$, for some $V\in\Rem{\cf^\phi}{U}$, then $x,W\sat\phi$, for
all $W\in\Rem{\cf^\phi}{U}$,  since  $\cf^\phi$ is stable for $\phi$. Therefore
$x,W\sat\dam\phi$ for all $W\in\Rem{\cf^\phi}{U}$.
If $x,V\sat\phi$, for some $V\sbs U$ with $V\not\in\Rem{\cf^\phi}{U}$, then  we
have  $V\sbs W$, for all $W\in\Rem{\cf^\phi}{U}$, since
the set of subsets containing $x$ is linearly ordered and $\Rem{\cf^\phi}{U}$
is stable and convex. Hence $x,W\sat\dam\phi$, for
all $W\in\Rem{\cf^\phi}{U}$.
\end{itemize}
\smallskip
\end{prf}

The following corollary is ``folklore''.

\begin{corol}
The formula  $\Box\dam\phi\to\dam\Box\phi$ is sound in  treelike spaces.
\label{corol:sound}
\end{corol}

\begin{prf}
Let $x,U\sat\Box\dam\phi$ in some model $\langle X,\co,i\rangle$.

By the Partition theorem, there exists a finite stable partition $\cf$ for
$\phi$. Further, there is a $V\in\cf$ which is ``the least''
in the following sense: if $W,W'\in\co$ contain $x$, $W\in\Rem{\cf}{V}$, and
$W'\sbs W$ then we will also have $W'\in\Rem{\cf}{V}$. The existence of such
a set $V$ is assured by the fact that
$\cf$ is finite, the members of the partition which
$\cf$ induces are convex, and the set of subsets in $\co$ which
contain $x$ is linearly ordered.
Moreover, $\Rem{\cf}{V}$ contains at least one subset which contains $x$, say
$W$.

Now, we have
 either $U\sbs V$ or $V\sbs U$. In the former case, we have $U\in\Rem{\cf}{V}$.
Hence $x,U\sat\Box\phi$ as $\Rem{\cf}{V}$ is stable for $\phi$. In the latter
case,
$x,W\sat\dam\phi$, since $W\sbs V \sbs U$. Thus we have
$x,W\sat\Box\phi$
for the same reasons as above. Hence $x,U\sat\dam\Box\phi$.
\end{prf}

A finite partition does not have a treelike form. Therefore we cannot perform a
filtration in a direct manner. First, we shall consider no partition member
(remainder) that contains no subset belonging to the initial treelike space.
Next, we shall impose a relation $\leq$ among the remaining members
(Definition~\ref{def:remainder-relation}). Two remainders will be related just in
case they contain subsets with common elements. This relation is not a partial
order. However, it respects the initial treelike ordering
(Lemma~\ref{lem:reduct} through \ref{lem:semitrans}). Finally, using a number of
equivalence relations based on $\leq$, one for each member of the partition, we
shall construct a treelike model equivalent to the initial one
(Propositions~\ref{prop:partition-is-treelike} and \ref{prop:equivalent-models}).
Moreover, the underlying space of this model will contain a finite number of
subsets.

By the Partition theorem, given a treelike  model $\langle X,\co,i\rangle$ and
a formula $\phi$, there exists a  finite partition $\cf^\phi$ on $\co$ stable for
$\phi$. For each $U\in \cf^\phi$, let
$$
\overline{U}\quad=\quad \bigcup\Rem{\cf^\phi}{U}
$$
and
$$
\overline{\cf^\phi}\quad = \quad\big\{ U \mid U \in \cf^\phi
\hbox{and} \ \overline{U}\not=\emptyset \big\}. $$

We have the following

\begin{lemma}
If ${U_1}, {U_2} \in \overline{\cf^{\phi}}$ with  $\overline{U_1} \subset
\overline{U_2}$, and $V_1, V_2\in \co$ with
$V_1 \in \Rem{\cf^\phi}{U_1}$, $V_2  \in \Rem{\cf^\phi}{U_2}$ and $V_1\cap
V_2 \not= \emptyset$, then $V_1 \subset V_2 $.
\end{lemma}

\begin{prf}
Since $V_1\cap V_2 \not= \emptyset$, then, by connectedness, we have either
$V_1\sbs V_2$ or $V_2\sbs V_1$. If $V_1\sbs V_2$ then $ V_1\subset V_2$ since
they belong to distinct equivalence classes.
If $V_2\sbs V_1$ then we have $ V_2\sbs V_1 \sbs \overline{U_1} \sbs
\overline{U_2}$. Hence $V_1 \in \Rem{\cf^\phi}{U_2}$, by
Proposition~\ref{prop:part}(\ref{convex}).
\end{prf}

\begin{dfn}\label{def:remainder-relation}
Let $<$ be the following relation on $\overline{\cf^\phi}$
$$
\begin{array}{lll}
U_1 < U_2  & \hbox{iff} &
                              \overline{U_1} \cap \overline{U_2}\not=\emptyset,\
                      \hbox{and}\\
               &           &        \hbox{for all}\ x,V_1,V_2 \ \hbox{such that}\
                          x \in\overline{U_1}\cap
                           \overline{U_2},\ V_1\in
                           \Rem{\cf^{\phi}}{U_1}\ \hbox{with}\ x\in V_1,\\
               &            & \hbox{and}\ V_2\in\Rem{\cf^{\phi}}{U_2}\
                           \hbox{with}\ x\in V_2, \  V_1\subset V_2.
\end{array}
$$

Clearly, we cannot have $U_1< U_2$ and $U_2 < U_1$.
Let $U_1 \leq U_2$, if either $U_1=U_2$ or $U_1 < U_2$.
\end{dfn}

The following lemma allows us to weaken the conditions of the definition of $<$.

\begin{lemma}\label{lem:reduct}
Let $U_1,U_2\in \overline{\cf^\phi}$ with $U_1\not= U_2$.  If there exist $x
\in\overline{U_1} \cap \overline{U_2}$ and $V_1\in \Rem{\cf^{\phi}}{U_1}$,
$V_2\in\Rem{\cf^{\phi}}{U_2}$
with $x\in V_1\cap V_2$ such that $V_1\subset V_2$, then $U_1<U_2$.
\end{lemma}

\begin{prf}
Suppose, towards a contradiction, that for
$y\in \overline{U_1}\cap\overline{U_2}$ there exist
$W_1\in\Rem{\cf^{\phi}}{U_1}$
and  $W_2\in \Rem{\cf^{\phi}}{U_2}$ such that $y\in W_1\cap W_2$ and
$W_2\sbs W_1$. By our hypothesis, $U_1\not= U_2$ and $V_1\subset V_2$, and so we
have $U_2\not\sbs U_1$.
This implies that $\Rem{\cf^{\phi}}{U_2}\cap \dar U_1 = \emptyset $. Therefore
$W_2 \not\in\Rem{\cf^{\phi}}{U_2}$ which is a contradiction. Thus
$W_1 \subset W_2$. Hence $U_1 < U_2$.
\end{prf}

\begin{lemma}\label{lem:utree}
Let $U_1,U_2\in \overline{\cf^\phi}$.
If $\overline{U_1}\cap\overline{U_2}\not= \emptyset$
then $U_1\leq U_2$ or $U_2 \leq U_1$.
\end{lemma}

\begin{prf}
Suppose that $U_1\not= U_2$ and let $x\in \overline{U_1}\cap\overline{U_2}$. Let
$V_1\in \Rem{\cf^{\phi}}{U_1}$
and  $V_2\in \Rem{\cf^{\phi}}{U_2}$ such that $x\in V_1\cap V_2$. Since $\co$
is a treelike space, we have either $V_1\sbs V_2$ or $V_2\sbs V_1$. Suppose that
the former holds. Since $U_1\not= U_2$, we have $V_1\subset V_2$. By
Lemma~\ref{lem:reduct}, $U_1< U_2$.
Similarly, if $V_2\sbs V_1$ then $U_2 < U_1$.
\end{prf}

\begin{lemma}\label{lem:antirefl}
$\leq$ is reflexive and antisymmetric.
\end{lemma}

\begin{prf}
Reflexivity  is straightforward.
For antisymmetry, suppose that $ \overline{U_1}\cap\overline{U_2}\not=
\emptyset$, $U_1\leq U_2$ and $U_2\leq U_1$. If $U_1\not= U_2$ then we
have $U_1< U_2$ and $U_2 < U_1$ which is a  contradiction.
\end{prf}

Instead of transitivity, we have the following property of $\leq$:

\begin{lemma}\label{lem:semitrans}
Let $U_1,U_2, U_3 \in \overline{\cf^\phi}$. If $U_1\leq U_2$, $U_2\leq U_3$ and
$\overline{U_1}\cap \overline{U_2} \cap \overline{U_3} \not=\emptyset$ then
$U_1\leq U_3$.
\end{lemma}

\begin{prf}
If either
$U_1= U_2$ or $U_2 =U_3$ we are done, so  suppose that $U_1<U_2$
and $U_2< U_3$. Let $x\in \overline{U_1}\cap \overline{U_2} \cap
\overline{U_3}$,
$V_1\in\Rem{\cf^{\phi}}{U_1}$ and $V_3\in\Rem{\cf^{\phi}}{U_3}$ such that
$x\in V_1$ and $x\in V_3$. Since $x\in \overline{U_2}$, there exists
$V_2 \in \Rem{\cf^{\phi}}{U_2}$ such that $x\in V_2$. Also, we have
$V_1 \subset V_2 \subset V_3$, since $U_1< U_2$ and
$U_2< U_3$. So, by
Lemma~\ref{lem:reduct},  $U_1< U_3$.
\end{prf}

Since $\overline{\cf^\phi} \sbs \cf^\phi$, $\overline{\cf^\phi}$ is finite.
Let $\overline{\cf^\phi}=\{ U_1, U_2, \ldots, U_n \}$, for some
$n$.
Now, let $\sim_i$ be the following equivalence relation on $\overline{U_i}$
$$
\begin{array}{lll}
x\sim_i y & \hbox{iff} & \hbox{for all}\ \overline{U_j}, j\in\{1,2,\ldots,n\}\
                  \hbox{such that}\ U_i\leq U_j,\\
          &      & x\in \overline{U_j}\ \hbox{iff}\ y\in \overline{U_j}.
\end{array}
$$

We denote the equivalence of $x$ under $\sim_i$ with $[x]_i$. Observe that the
number of equivalence classes is finite, since it depends only on the number of
members of the partition.

\begin{lemma} \label{lem:simless}
\renewcommand{\theenumi}{\alph{enumi}}

Let $U_k,U_l \in \overline{\cf^\phi}$, $k,l\in \{1,2, \ldots,n \}$, with
$ \overline{U_k} \cap \overline{U_l} \not= \emptyset$.
Then
\begin{enumerate}

\item if $U_k \leq U_l$ then $[x]_k \sbs [x]_l$ ,
for all $x\in \overline{U_k}\cap \overline{U_l}$, and \label{part-one}

\item if $[x]_k \subset [x]_l$, for some $x\in \overline{U_k}\cap
\overline{U_l}$, then $U_k < U_l $. \label{part-two}

\end{enumerate}

\end{lemma}

\begin{prf}
For Part~\ref{part-one}, if $U_k=U_l$ then we are done. Suppose  $U_k < U_l$
and let $z \in [x]_k $. Let $U_m\in \overline{\cf^\phi}$, $m\in
\{1,2,\ldots,n\}$, be such that $U_l \leq U_m$. If $x\in \overline{U_m}$ then
$x\in \overline{U_k} \cap \overline{U_l} \cap \overline{U_m}$. So, by
Lemma~\ref{lem:semitrans}, $U_k \leq U_m$. So $z \in \overline{U_m}$,
since $x\sim_i z$. For the other direction, suppose $z \in \overline{U_m}$. Then
we have  $z \in \overline{U_l}$,
since $U_k \leq U_l$ and $x \sim_1 z$. So
$z \in \overline{U_k} \cap \overline{U_l} \cap \overline{U_m}$.
Hence, by Lemma~\ref{lem:semitrans}, $U_k \leq U_m$. Also, $x \in
\overline{U_m}$, since $x\sim_i z$. Therefore $z \in [x]_l$.

For Part~\ref{part-two}, we have either $U_k < U_l $ or $ U_l \leq U_k$, since
$\overline{U_k}\cap \overline{U_l} \not= \emptyset$. Suppose the latter towards
a contradiction.  Then, by Part~\ref{part-one} and Lemma~\ref{lem:utree},
we have  $[x]_l\sbs [x]_k$ which is a contradiction to our hypothesis.
\end{prf}

\begin{lemma} \label{lem:simlessb}
Let $U_k,U_l \in \overline{\cf^\phi}$, $k,l\in \{1,2, \ldots,n \}$, with
$ \overline{U_k} \cap \overline{U_l} \not= \emptyset$.
If $U_k < U_l$ then $[x]_k \subset [x]_l$, for all $x\in \overline{U_k}\cap
\overline{U_l}$.
\end{lemma}

\begin{prf}
By Lemma~\ref{lem:simless}(\ref{part-one}), we have $[x]_k \sbs [x]_l$. Suppose
$[x]_k = [x]_l$. Let $V\in \Rem{\cf^\phi}{U_i}$ such that $x\in V$.
We have $V\sbs [x]_i$. Thus $V\sbs [x]_j \sbs \overline{U_j}$.
So, for each $y\in V$, there exists $V_y \in \Rem{\cf^\phi}{U_j}$ such that
$V_y\subset V$. But then $V=\bigcup_{y\in V}V_y \in \Rem{\cf^\phi}{U_j}$ which is
a contradiction, since $U_i\not= U_j$.
\end{prf}

Now, let
$$[\overline{\cf^\phi}]=\{ [x]_i \mid x\in\overline{U_i},\ i\in\{1,2,\ldots,n
\}\}.$$

\begin{prop}\label{prop:partition-is-treelike}
The subset space $\langle X, [\overline{\cf^\phi}] \rangle$ is a treelike space.
\end{prop}

\begin{prf}
First notice that $ X \in \overline{\cf^\phi} $, since $ X \in \cf^\phi$.
Thus $X=U_{i_0}$, for some $i_0\in \{1,2,\ldots,n\}$. Moreover, $x\sim_{i_0} y$,
for all $x,y \in X$. Hence $ X = [x]_{i_0} \in [\overline{\cf^\phi}]$.

Now, let $[x]_i \cap [y]_j \not=\emptyset$, for some $x\in \overline{U_i}$ and
$y \in \overline{U_j}$. Let $z\in [x]_i \cap [y]_j$. We have
$[x]_i=[z]_i$ and $[y]_j= [z]_j$. Further, $z\in \overline{U_i}$ and
$z\in \overline{U_j}$, i.e. $ \overline{U_i} \cap \overline{U_j} \not=
\emptyset $.
So, by Lemma~\ref{lem:utree}, we have either $U_i \leq U_j$ or $U_j
\leq U_i$. By Lemma~\ref{lem:simless}(a), we have either
$[z]_i \sbs [z]_j$ or $[z]_j \sbs [z]_i$, respectively. Therefore
either $[x]_i \sbs [y]_j$ or $[y]_i \sbs [x]_j$.
\end{prf}

Let $\overline{\cm}=\langle X, [\overline{\cf^\phi}],\overline{i} \rangle$ be
the treelike  model where $\overline{i}(A) = \{ [x]_i \mid x \in i(A) \}$.

\begin{prop}
\label{prop:equivalent-models}
For all $x\in X$, $V\in \co$ and $\psi\in\cl$ such that $\psi$ is a subformula
of $\phi$, if $V\in \Rem{\cf^{\phi}}{U_i}$, for some $i\in\{ 1,2,\ldots,n \}$,
then
$$
x,V\msat\psi \qquad \hbox{iff} \qquad x,[x]_i\sat_{\overline{\cm}} \psi.
$$
\end{prop}

\begin{prf}
By induction on the complexity of $\phi$. The only interesting case is that of
$\phi=\Box\psi$. Suppose $x,[x]_i\sat\dam\neg\psi$ but $x,V\msat\Box\psi$,
for some $V \in \Rem{\cf^\psi}{U_i}$. The latter implies
that there is $j\in \{1,2,\ldots,n\}$ such that $x,[x]_j\sat\neg\psi$ and
$[x]_j\sbs [x]_i$. We have $x,[x]_i\sat\psi$, by  $x,V\msat\psi$ and induction
hypothesis. Hence $[x]_j\subset [x]_i$. By Lemma~\ref{lem:simless}(b), we have
$U_j < U_i$. By induction hypothesis, we have $x, V'
\msat \neg \psi$, for all $V' \in \Rem{\cf^\psi}{U_j}$ such that $x\in V'$.
Also, we  have $V' \subset V$, since   $\overline{U_j}<\overline{U_i}$. Hence
 $x,V\msat \neg \psi$,  a contradiction.

Now, suppose $x, [x]_i \sat\Box \psi$ but $x,V\msat\dam\neg\psi$, for some
$V\in \Rem{\cf^\psi}{U_i}$. So there exists
$V' \in \Rem{\cf^\psi}{U_j}$ such
that $x\in V'$, $V'\sbs V$, and $x,V'\msat \neg \psi$. Also,
$x,[x]_i\sat\psi$ so, by induction hypothesis, $x,V\msat\psi$. The latter implies
 $U_i\not= U_j$, since $\Rem{\cf^\psi}{U_i}$ is stable for $\psi$. Therefore
we have $U_j < U_i$. Hence, by Lemma~\ref{lem:simlessb},
$[x]_j\subset [x]_i$. Thus  $x, [x]_j\sat\neg\psi$, by induction hypothesis.
Hence
$x,[x]_i\sat\dam\neg\psi$, a contradiction to our hypothesis.

\end{prf}

Constructing the above model is not adequate
for generating a finite model, since there may still be an infinite
number of points. It turns out that we only need a finite number of
them.

Let $\cm=\langle X,\co,i\/\rangle$ be a
treelike  model, and define an equivalence relation $\sim$ on $X$ by
$x\sim y$ iff

\begin{enumerate}
\item for all $U \in \co$, $x\in U$ iff $y \in U$, and
\item for all atomic $A$, $x\in i(A)$ iff $y \in i(A)$.
\end{enumerate}

Further, denote by $x^*$ the equivalence class of $x$,
and let $X^* = \{x^* : x \in X\}$. For every $U\in\co$, let  $U^*=\{x^*
: x\in U\}$, then $\co^* = \{ U^* : U \in \co \}$ is a treelike space on
$X^*$. Define a map $i^*$ from the atomic formulae to the powerset
of $X^*$ by $ i^*(A) = \{ x^* : x\in i(A) \} $. The entire model $\cm$
lifts to the model $\cm^* = \langle X^*,\co^*, i^* \/\rangle $ in a
well-defined way.

\begin{lemma} For all $x$, $U$, and $\phi$,
$$x, U \sat_\cm \phi \qquad\mbox{iff} \qquad x^*, U^* \sat_{\cm^*} \phi\ .$$
\label{lemma:quotient}
\end{lemma}

\begin{prf}
By induction on $\phi$.
\end{prf}

\begin{theorem}
If $\phi$ is satisfied in any treelike   space then $\phi$
is satisfied in a finite treelike space. \label{thm:finiteness}

\end{theorem}

\begin{prf}
Let $\cm=\langle X,\co,i\/\rangle$ be such that, for some $x\in U \in
\co$, $x, U \sat_\cm \phi$.
Let $\cf^{\phi}$ be a finite stable partition (by
Theorem~\ref{thm:main})
for $\phi$ and its subformulae with respect to $\cm$.
By Proposition~\ref{prop:equivalent-models},
$x, U \sat_\cn \phi$, where $\cn=\langle X,\cf,i\/\rangle$. We may
assume that $\cf$ is a treelike space,
and we may also assume that the overall language has only the
(finitely many) atomic symbols which occur in $\phi$.
Then the relation $\sim$ has only finitely many classes.
So the model $\cn^*$ is finite. Finally, by
Lemma~\ref{lemma:quotient}, $x^*, U^* \sat_{\cn^*}\phi$.
\end{prf}

Observe that  the finite
treelike space
is a quotient of the initial one under two equivalences. The one equivalence is
on the elements of the treelike space and the number of
equivalence classes is  a function of the complexity of $\phi$. The other
equivalence is  on the points of the treelike space and
the number of equivalence classes is a function of the atomic formulae
appearing in $\phi$. So the overall size of the (finite) treelike
space is bounded by a function of the complexity of $\phi$. Thus if we
want to test if a given formula is invalid we have a finite number of
finite treelike spaces where we have to test its validity. Thus we
have the following

\begin{theorem}
The theory of treelike spaces is decidable.
\end{theorem}
\medskip

\noindent {\bf Acknowledgments:} The author is indebted to Rohit Parikh for
bringing this problem to his attention and wishes to thank Bernhard
Heinemann, Larry Moss, and Timothy Williamson, as well as, the anonymous referees
for helpful comments.

\newcommand{\etalchar}[1]{$^{#1}$}

\end{document}